\documentclass{article}

\usepackage{arxiv}

\usepackage{algorithm}
\usepackage{algpseudocode}
\usepackage[utf8]{inputenc} 
\usepackage{url}            
\usepackage{graphicx}
\usepackage{amssymb}
\usepackage{amsmath}
\usepackage{multirow}
\usepackage{hyperref}       
\usepackage{adjustbox}

\graphicspath{ {./images/} {./} }

\title{Efficient bandwidth extension of musical signals using a differentiable harmonic plus noise model}

\author{ \hspace{1mm}Pierre-Amaury Grumiaux\\
	Nantes Universit{\'e}, {\'E}cole Centrale Nantes, CNRS, LS2N\\
	2 chemin de la Houssini\`ere\\
	F-44332 Nantes, France \\
	\texttt{pierreamaury.grumiaux@gmail.com}
	\And
	\hspace{1mm}Mathieu Lagrange \\
	Nantes Universit{\'e}, {\'E}cole Centrale Nantes, CNRS, LS2N\\
	2 chemin de la Houssini\`ere\\
	F-44332 Nantes, France \\
	\texttt{mathieu.lagrange@ls2n.fr}
}

\date{}

\renewcommand{\headeright}{}
\renewcommand{\undertitle}{}

\hypersetup{
pdftitle={Efficient BWE of musical signals using DDSP},
pdfsubject={cs.SD},
pdfauthor={P.A.~Grumiaux, M.~Lagrange},
pdfkeywords={},
}

\begin{document}
\maketitle

\begin{abstract}
The task of bandwidth extension addresses the generation of missing high frequencies of audio signals based on knowledge of the low-frequency part of the sound. This task applies to various problems, such as audio coding or audio restoration. In this article, we focus on efficient bandwidth extension of monophonic and polyphonic musical signals using a differentiable digital signal processing (DDSP) model. Such a model is composed of a neural network part with relatively few parameters trained to infer the parameters of a differentiable digital signal processing model, which efficiently generates the output full-band audio signal. 

We first address bandwidth extension of monophonic signals, and then propose two methods to explicitely handle polyphonic signals. The benefits of the proposed models are first demonstrated on monophonic and polyphonic synthetic data against a baseline and a deep-learning-based resnet model. The models are next evaluated on recorded monophonic and polyphonic data, for a wide variety of instruments and musical genres. We show that all proposed models surpass a higher complexity deep learning model for an objective metric computed in the frequency domain. A MUSHRA listening test confirms the superiority of the proposed approach in terms of perceptual quality.
\end{abstract}


\section{Introduction}
Audio bandwidth extension (BWE) is a subtask of audio enhancement \cite{vincent_audio_2018} whose goal is to extrapolate the audio spectrum to higher frequencies, in contrast with \textit{audio inpainting} whose goal is to interpolate missing parts \cite{adler_audio_2012}. 
BWE has been considered early in telecommunication systems to overcome bandwidth limitations, especially in telephony for which the typical sampling rate is $16$~kHz, \emph{i.e.}, leading to the highest frequency in the processed signal be $8$~kHz. In the case of human conversations, the quality of speech can be greatly improved if the sampling rate is increased to $44.1$ or $48$~kHz \cite{french_factors_1947}. In the same vein, another application of BWE is to improve the quality of old music recordings, possibly in addition to the removal of clicks and noise \cite{vaseghi_restoration_1992} or declipping \cite{gaultier_sparsity-based_2021}. In both applications, the signal enhancement is handled without access to the original signal with better quality. Informed BWE algorithms can also be useful in audio coding \cite{dietz_spectral_2002} where signals of smaller sampling rates are more effectively compressed, requiring the use of a BWE to restore the full sampling rate of the decoded signal. In most cases low bitrate side information is transmitted along the compressed low frequency signal to improve the performance of the BWE module.

Finally, BWE is also meaningful for interoperability of audio processing tools as many audio signal processing methods, such as source separation \cite{vincent_audio_2018}, speech synthesis \cite{ning_review_2019} or voice conversion \cite{mohammadi_overview_2017}, focus on $16$-kHz signals, hence the need for a BWE system beforehand if the acquired signal is not at the desired sampling rate.

Even-though many deep learning based systems have been proposed to tackle BWE, most of them do not consider runtime efficiency as critical, leading to high quality systems that can be very costly at inference. High quality generators based on autoregressive signal models such as Wavenet \cite{oord2016wavenet} or on diffusion \cite{moliner_solving_2023} have intrinsic high complexity and sequentiality which limit their use for time or delay critical applications. 

In this paper, we propose to consider differentiable digital signal processing (DDSP) models derived from the seminal work of \cite{engel_ddsp_2020} in order to tackle BWE in an efficient manner.  
Controlling an harmonic plus noise sound model with a deep learning architecture allows us to considerably reduce inference time. Experiments described in this paper demonstrate a speed increase a 100 \% compared to a reference resnet implementation \cite{sulun_filter_2021}, this with better resulting perceptual quality. This is due to several factors, including reduction of learnable parameters. Using the DDSP approach, the sound is generated using deterministic synthesizers that are controlled by several deep-learning modules of relatively small sizes. For comparison, the resnet architecture has more than 55k learnable parameters, while the tested DDSP approach has around 4k parameters.

The remaining of this article is organized as follows. In Section~\ref{sec:related_works}, we present a general overview of works related to BWE. In Section~\ref{sec:models}, we explain the proposed models designed to address BWE. The experimental protocol whose code is available online\footnote{ \url{https://github.com/mathieulagrange/ddspMusicBandwidthExtension}.},  and which rely on publicly avaible datasets, is detailed in Section~\ref{sec:method}. In Section~\ref{sec:synthetic_data}, we show how the proposed models are well designed when considering synthetic data, and in Section~\ref{sec:evaluation} they are evaluated on real data. Finally, we conclude this article in Section~\ref{sec:conclusion}.

\section{Related work}
\label{sec:related_works}

Most approaches considered speech signals with application to telephony. The literature that consider music or general audio is more scarce. When available, we here put the focus on literature related to musical audio.

\subsection{Signal processing approaches}

Early works employed pure signal processing methods for BWE. In the area of audio coders, some non-blind systems rely on spectral band replication (SBR) \cite{dietz_spectral_2002} using side information extracted during compression. The SBR algorithm is based on the replication of the low-band spectrum to the high-band region, possibly with the benefits of side information about the high frequencies to improve the overall performance. It has been extended in several works \cite{meltzer_sbr_2002, nagel_harmonic_2009}, \emph{e.g.}, by replacing the replication by a stretching of the low-band content towards the high-band part, thus preserving the intrinsic harmonic relationships. Source-filter models have been also employed to extend the bandwidth using line spectral frequencies in \cite{chennoukh_speech_2001}. Systems based on dictionary learning to map low-frequency patterns to high-frequency components has been proposed in \cite{sadasivan_joint_2016, yoshida_algorithm_1994}. Classic machine learning methods have also been explored for BWE, such as Gaussian mixture models (GMMs) \cite{park_narrowband_2000}, hidden Markov models (HMM) \cite{bauer_hmm-based_2008, song_study_2009} or non-negative matrix factorization (NMF) \cite{bansal_bandwidth_2005, sun_non-negative_2013}.

\subsection{Convolutional deep learning approaches}

Recently, deep learning (DL) methods have shown great performance to synthesize the upper band spectrum. The first works that apply DL technique in BWE literature used deep neural networks (DNNs) with dense layers to infer the high frequencies up to $8$~kHz \cite{li_deep_2015, li_dnn-based_2015, wang_speech_2015}. In \cite{li_deep_2015}, the log short-time Fourier transform is fed into several dense layers with the last one inferring the high-band spectrum magnitude. The waveform is reconstructed by using the flipped phase from the low-band to estimate the high-band phase information. While this flipped method avoids having phase discontinuities at the low/high frontier, \cite{li_dnn-based_2015} propose to cope with this potential issue by extended the mean-squared error (MSE) loss function with a regularization term. Gaussian-Bernoulli restricted Boltzmann machines (GBRBM) has been employed alongside dense layers in \cite{wang_speech_2015} in order to estimate the higher spectral envelope. Other systems make use of convolutional neural networks (CNNs) to infer the high frequencies from the low-band input features, using 1D convolutions in the time domain \cite{gu_waveform_2017, kuleshov_audio_2017, wang_time-frequency_2020, sulun_filter_2021} or 2D convolutions in the spectro/temporal domain \cite{campos_high_2018, lagrange_bandwidth_2020}. In \cite{gu_waveform_2017}, the authors show that using a network architecture of 1D dilated convolutions and residual connections outperforms a state-of-the-art based on a long short-term memory (LSTM) system on speech signals. The authors of \cite{kuleshov_audio_2017} make use of 1D convolutional layers in an encoder-decoder scheme to extend the bandwidth of speech and musical signals in three upscaling ratios: $2$, $4$ and $6$. They show the effectiveness of their system compared to \cite{li_dnn-based_2015} for objective and perceptive metrics. In the same vein, an encoder-decoder architecture in the time domain is also used in \cite{wang_time-frequency_2020}, but the authors propose to opt for subpixel layers instead of classical transposed convolutional layers because it was shown that less artifacts are created by considering those layers. Another encoder-decoder system can be found in \cite{sulun_filter_2021} where the architecture also contains residual connections in a U-Net scheme. The authors shows 1) that using a Resnet architecture outperforms the U-Net, probably because of the loss of information in the bottleneck layer of the former, and 2). that the employed DNNs overfit on the filter shapes present in the training data. This latter problem can fortunately be alleviated with a data augmentation strategy which utilizes a wide variety of low pass filters during training. 

As the paper describes thoroughly the architecture proposed as well as its learning procedure, we choose to use this latter system as a reference "high complexity" system.

\subsection{Generative adversarial networks}
Generative adversarial networks (GAN) have been explored in several works for BWE. In \cite{li_speech_2018}, the authors show that relying on GANs can improve the generated speech quality by using a simple DNN. In \cite{li_speech_2019}, the generator is based on a U-Net-like architecture and the discriminator is trained to distinguish between generated and true wide-band signals, with the addition of a perceptual loss expressing the distance between features learned by a pre-trained automatic speech recognition (ASR) network. A combination of two discriminators, one based on spectral features and the other based on temporal features, have been proposed in \cite{su_bandwidth_2021} to extend the bandwidth from $8$~kHz to $48$~kHz. In \cite{moliner_behm-gan_2022}, the generator is also based on a U-Net architecture yet it is proposed to employ CNNs for the three discriminators, each one being applied on a downsampled version of generated or true waveform (downsampling factors = $1$, $2$, $4$). The generator is then trained to generate piano signals. 

While GAN do not impose strong constraints in terms of inference complexity, GANs are known to be notoriously difficult to train, as they require very specific choices in optimization and architectures in order to stabilize training and could fail to cover modes of the data distribution \cite{song2020denoising}. 

At the time of the design of this study, we found no pretrained general audio BWE model learnt with an adversarial procedure. We thus do not consider a GAN trained generator as another reference method. 

\subsection{Diffusion models}

In terms of quality of generation, diffusion models now provides very convincing performance for a wide variety of data, including audio \cite{moliner_solving_2023, moliner2023zero}. As for autoregressive architecture like Wavenet \cite{oord2016wavenet}, this important increase of quality comes at a strong computational cost at inference. The network has to be called sequentially a large number of times (usually from 100 to 1000 times) in order to perform the inference. One can reduce the size of the network or reduce the number of steps in order to accelerate sampling \cite{song2020denoising}, but those approaches are detrimental to the quality of the generated audio and the inference time remains high.

In this paper, we find that the inference of a standard ResNet architecture is already about 1000 times real time on a standard central processing unit (CPU) and our study focuses on efficient BWE, we choose not to consider diffusion models as a reference. See Figure \ref{fig:lsd_vs_time} for more details about computation.

\section{Differentiable sound models}
\label{sec:models}
In this article, we address BWE using DDSP models derived from the seminal work of \cite{engel_ddsp_2020} that focuses on the generation of audio signals with a combination of neural networks and digital signal processing models. This approach allows one to train the neural network parameters in an end-to-end fashion with backpropagation, if the rest of the model is differentiable. Besides several sound synthesis models \cite{hayes_neural_2021, shan_differentiable_2021}, DDSP has also been successfully applied to other tasks, such as neural audio effect \cite{lee_differentiable_2022}, style transfer \cite{steinmetz_style_2022}, sound matching \cite{masuda_synthesizer_2021} or virtual analog \cite{esqueda_differentiable_2021}.

In this section, we describe the DDSP models we propose for monophonic and polyphonic BWE.

\subsection{Monophonic BWE system}

\begin{figure*}
\centering
\includegraphics[width=1.\columnwidth]{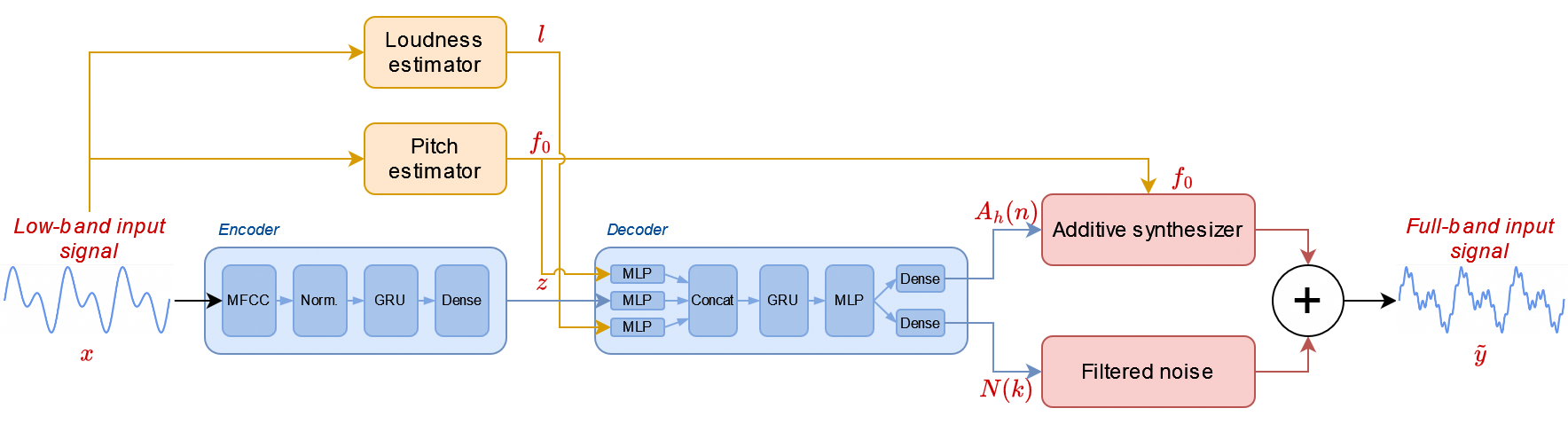}
\caption{DDSP architecture for monophonic bandwidth extension (model \textit{DDSP-mono-dec}).}
\label{fig:ddsp_mono}
\end{figure*}

To address BWE for monophonic musical signal, we adapt the model proposed in \cite{engel_ddsp_2020}, which is monophonic by design. The main difference with the original DDSP model is that, in order to reconstruct the higher frequencies, the model takes as input the low-band (LB) audio signal of bandwidth $\frac{f_N}{\alpha}$, with $f_N$ the Nyquist frequency, and is trained to output the wide-band (WB) signal of bandwidth $f_N$. The overall architecture, illustrated on Fig.~\ref{fig:ddsp_mono} is the same as in \cite{engel_ddsp_2020}, and consists in two parts: an trainable encoder-decoder neural network, and a harmonic-plus-noise synthesizer. The neural network is illustrated in blue, the extracted features are shown in yellow, and the differentiable synthesizer is colored in red. This monophonic model is labeled \textit{DDSP-mono-dec}, referring to the design of the decoder to generate monophonic parameters.

\subsubsection{Extracted features}

The input LB signal is first analyzed to extract the fundamental frequency $f_0(n)$ and loudness $l(n)$ over time. In the monophonic setting, we use CREPE \cite{kim_crepe_2018}, a state-of-the-art monophonic pitch estimator based on a convolutional neural network, to estimate $f_0$. The loudness $l$ is obtained with a A-weighting of the power spectrum \cite{hantrakul_fast_2019}.

\subsubsection{Neural network}
\label{sec:ddsp_mono_ae}
 The input LB signal waveform is processed by an encoder which creates an latent vector $z$. In the encoder, the first 30 mel frequency cepstrum coefficients (MFCC) are extracted from the audio input (fast Fourier transform (FFT) size of 1024, overlap of $75$~\% and 128 mels between $20$~Hz and $8000$~Hz) and then passed into a trainable normalization layer. After that, the MFCCs goes into a gated recurrent unit (GRU) with $512$ units and finally a $512$-neuron linear layer outputs the latent vector $z(n)$.

The three vectors $z(n)$, $f_0(n)$ and $l(n)$ are then fed into the decoder. Each of them first goes into a separate multi-layer perceptron (MLP) with three layers, and the outputs are concatenated. The obtained vector is processed by a 512-unit GRU and then a another 3-layer MLP. Finally, two separate dense layers are used : the first one outputs the harmonic amplitudes $A_h(n)$ (see Section~\ref{sec:harmonic_noise_synth}) using a softmax activation, and the second one gives the noise filter coefficients $N(k)$. Note that, as in \cite{engel_ddsp_2020}, we use a modified sigmoid function $\sigma(x)$ at the output of these two last dense layers : $\sigma(x) = 2 \cdot sigmoid(x)^{log(10)} + 10^{-7}$. This architecture has around 3k learnable parameters.

\subsubsection{Harmonic-plus-noise synthesizer}
\label{sec:harmonic_noise_synth}
 Both outputs from the neural network are used separately in the additive synthesizer and noise modules. The additive synthesizer takes the estimated $f_0(n)$ and the inferred harmonic amplitudes $A_h(n)$ to generate the audio signal $y(n)$:
 \begin{equation}
      y(n) = \sum_{h=1}^{H} A_h(n) sin(\phi_h(n)),
 \end{equation}
 
where $\phi_h$ is the instantaneous phase of the $h$-th sinusoidal component. It is computed by integrating the instantaneous frequency $f_h(n) = h f_0(n)$ : 

\begin{equation}
    \phi_h(n) = 2 \pi \sum_{m=0}^n f_h(m) + \phi_{0,h},
\end{equation}

where $\phi_{0,h}$ is a random initial phase.
In the filtered noise module, we obtain a time-domain finite impulse response (FIR )filter as the inverse discrete Fourier transform of the noise filter coefficients $N(k)$ from the neural network output. The filtered noise signal is synthesized by convolving a white noise with the FIR filter. The harmonic signal and filtered noise are finally summed to obtain the wide-band output signal.

Even if the full-band output signal is generated, only the missing high frequency content is kept, and added to the input low-band signal.

\subsubsection{Noise-only synthesizer}

We also consider a noise only synthesizer in which the output of the autoencoder only contains the noise filter coefficients $N(k)$. We label this model \textit{DDSP-noise}. This will allow us in the experimental part to evaluate the respective value of the harmonic and noise parts of the synthesizer.

\subsubsection{Loss function}

We use the multi-scale spectral (MSS) loss function to train our models computed on the missing high-frequency region. It is defined as $\mathcal{L}(y, \tilde{y}) = \sum_{s} L_s(Y_s, \tilde{Y}_s)$, where $Y_s$ and $\tilde{Y}_s$ are the high frequency magnitude spectrograms of the ground-truth signal $y$ and the reconstructed signal $\tilde{y}$, respectively, computed using a FFT size $s$, and :

\begin{equation}
    L_s(Y_s, \tilde{Y}_s) = ||Y_s - \tilde{Y}_s ||_1 + ||log Y_s - log \tilde{Y}_s ||_1,
\end{equation}

$|| \cdot ||$ being the common $L_1$ norm. Indeed, experiments demonstrated that it is preferable to compute each loss only on the high frequency region for solving the BWE task. We use the same set of FFT sizes as in \cite{engel_ddsp_2020}, that is $[2048, 1024, 512, 256, 128, 64]$ samples.

\subsection{Polyphonic BWE methods}

By the use of a single harmonic synthesizer, the DDSP architecture can only generate high frequency content harmonically from a single $f_0$. To address BWE for polyphonic musical signals, we propose two systems: a cyclic use of a monophonic BWE system detailed above, and a BWE system based on a polyphonic DDSP architecture. 

\subsubsection{Cyclic monophonic decoder}

\begin{figure*}[ht!]
\centering
\includegraphics[width=1.\columnwidth]{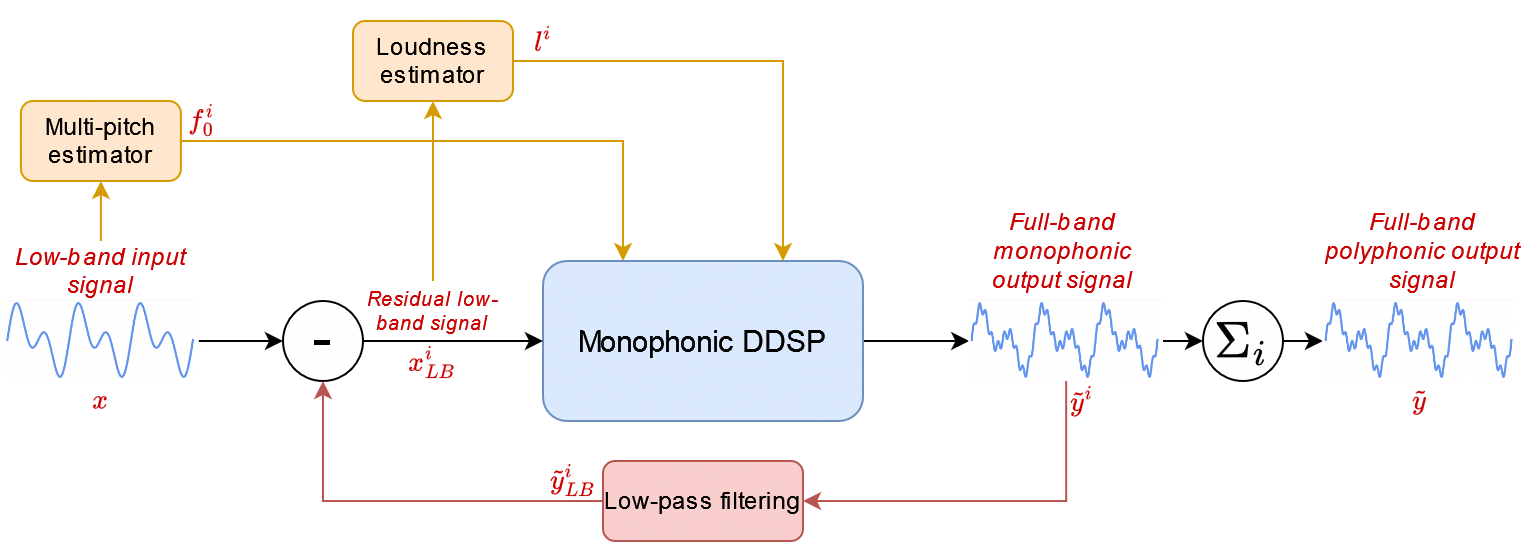}
\caption{Cyclic use of the monophonic DDSP model (\textit{DDSP-mono-dec-cyclic}) for bandwidth extension of polyphonic signals.}
\label{fig:ddsp_cyclic}
\end{figure*}

In the monophonic BWE system, the DDSP model generates an harmonic signal based on a single $f_0$ estimated from a monophonic pitch estimator \cite{kim_crepe_2018}. Now that we are in a polyphonic context, we use a state-of-the-art multi-pitch estimator \cite{bittner_lightweight_2022} which outputs a maximum of $I$ different fundamental frequencies $f_0^i$. Considering that this multi-pitch estimator has a rather good performance, we propose to iteratively use the monophonic DDSP model \textit{DDSP-mono-dec} in a cyclic manner, as illustrated in Fig.~\ref{fig:ddsp_cyclic}. We label this model \textit{DDSP-mono-dec-cyclic}. Pseudocode of the overall algorithm is detailed in Algorithm \ref{alg:ddsp_cyclic}.

\begin{algorithm}
\caption{Pseudocode algorithm of cyclic use of the monophonic DDSP model.}
\label{alg:ddsp_cyclic}
\begin{algorithmic}

\State $f_0^1, ..., f_0^I = multi\_pitch\_estimator(x)$
\State $x_{LB}^1 \gets x$
\State $\tilde{y} = 0$

\For{i := 1 to I step 1}
    \State $l^i = loudness\_estimator(x_{LB}^i)$
    \State $\tilde{y}^i = DDSP\_mono\_dec(x_{LB}^i, f_0^i, l^i)$
    \State $\tilde{y}_{LB}^i = low\_pass(\tilde{y}^i)$
    \State $\tilde{Y}_{LB}^i = STFT(\tilde{y}_{LB}^i)$
    \State $\tilde{X}_{LB}^i = STFT(\tilde{x}_{LB}^i)$
    \State $\lvert X_{LB}^{i+1} \rvert = \lvert X_{LB}^i \rvert - \lvert \tilde{Y}_{LB}^i \rvert$ \Comment{removing low-band content}
    \State $x_{LB}^{i+1} = \lvert X_{LB}^{i+1} \rvert e^{j \angle{X_{LB}^{i+1}}}$ \Comment{back to time domain}
    \State $\tilde{y} = \tilde{y} + \tilde{y}^i$ \Comment{constructing final signal step by step}
\EndFor

\State \textbf{return} $\tilde{y}$

\end{algorithmic}
\end{algorithm}

The monophonic DDSP model is applied for $I$ iterations on a low-band signal $x_{LB}^i$ which correspond to the original low-band signal minus the $i-1$ estimated sources. At each iteration $i$, a loudness contour $l^i$ is extracted from what we label a \textit{residual} low-band input signal $x_{LB}^i$ and passed, along with the $i^{th}$ estimated pitch $f_0^i$ (obtained on $x$ at the beginning of the algorithm) and $x_{LB}^i$, into the DDSP model.

The output full-band monophonic signal $\tilde{y}^i$, which contains a harmonic content from the current $f_0^i$, is then low-pass filtered to keep only the low-frequency part $\tilde{y}_{LB}^i$. Finally, the magnitude spectrogram of $\tilde{y}_{LB}^i$ is subtracted to the magnitude spectrogram of the residual low-band input signal:

\begin{align}
|X_{LB}^i| = \begin{cases} |X| & \text{if $i = 1$} \\
|X_{LB}^{i-1}| - |\tilde{Y}_{LB}^{i-1}| & \text{if $i \in \{2, ..., I\}$} \end{cases}
\end{align}

The low-band input signal is then obtained in the time-domain using an inverse short-term Fourier transform (STFT) on $X_{LB}^i$ (phase is kept in place).

In that way, at each iteration $i$, the harmonic content generated at the previous step is removed in the spectral domain from the residual low-band input signal, so that a different $f_0^i$ should be extracted. The residual low-band signal should contain less and less harmonics during this process.

At the beginning of the iteration, the loudness contour is then estimated on the full polyphonic signal, which will lead to estimations errors, that hopefully will decrease at each iteration.

The output of the noise synthesizer, which is part of the monophonic DDSP model at each iteration in order to have a more precise estimate of the amplitude of the harmonic of the sinusoidal part. While the noise part is thus estimated at each iteration we only considered the noise part of the last iteration $I$ in order not to overestimate the noise part. 

Finally, the full-band monophonic output signals $\tilde{y}^i$ are summed and mixed with the noise part to obtain the estimated full-band polyphonic signal $\tilde{y}$. As in the monophonic BWE setting, the high frequency content from this full-band signal is mixed with the low-band input signal.

\subsubsection{Polyphonic decoder}

\begin{figure*}[ht!]
\centering
\includegraphics[width=1.\columnwidth]{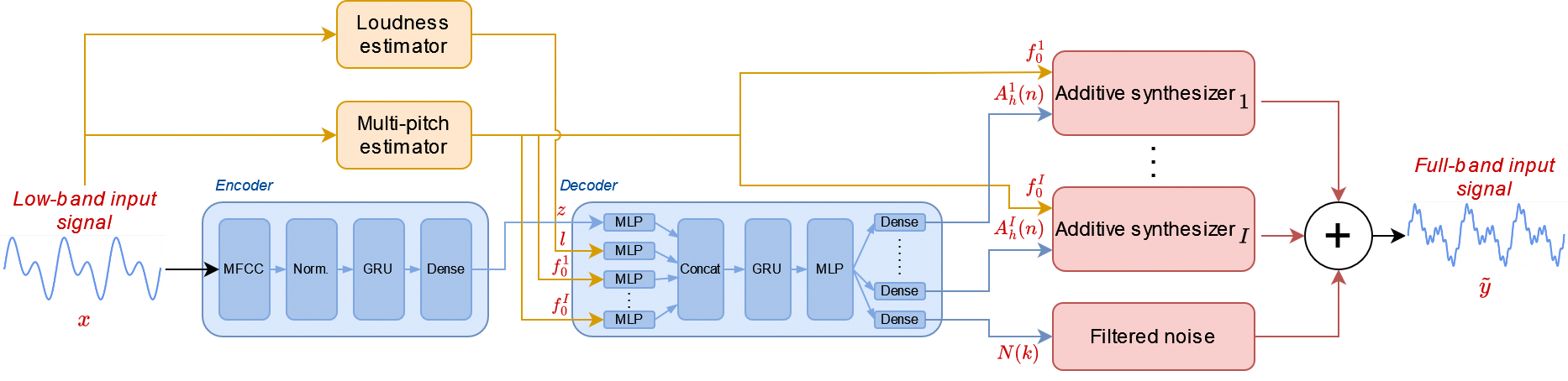}
\caption{Polyphonic DDSP model (\textit{DDSP-poly-dec}) for bandwidth extension.}
\label{fig:ddsp_poly}
\end{figure*}

To address BWE for polyphonic signals, we propose another model adapted from the original DDSP models, illustrated in Fig.~\ref{fig:ddsp_poly}, which we label \textit{DDSP-poly-dec} because the decoder outputs the parameters intended to control a polyphonic synthesizer. As before, the model is trained on polyphonic data.

In this model, $I$ additive harmonic synthesizers are used, where $I$ is the estimated number of fundamental frequencies $f_0^i, i \in {1, ..., I}$ present in the input low-band signal. To estimate the parameters for each separate additive synthesizer, we extend the decoder detailed in section \ref{sec:ddsp_mono_ae} by using $I$ separate MLPs for each $f_0^i$ (instead of a single MLP for vector $f_0$ in the monophonic DDSP model). The outputs of those $I$ MLPs are then concatenated into one vector, which is itself concatenated to the outputs of the two other MLPs applied on $z$ and $l$. Then, as in the monophonic model, the obtained vector goes through a GRU and another MLP. After that, $I+1$ dense layers are used: one for estimating the noise filter coefficients $N(k)$, and $I$ other layers to output the $H$ harmonic amplitudes of the $I$ additive synthesizers.

In this model, we employ the same multi-pitch estimator \cite{bittner_lightweight_2022} as in the cyclic model to estimate a maximum of $I$ $f_0^i$. If only $I' < I$ fundamental frequencies are given by the estimator, we set $f_0^i = 0, i > I'$, and all $f_0^i$ are fed in the decoder. To prevent any adverse impact on sound quality of those missing values, only the $I'$ first sets of $H$ harmonic amplitudes are extracted from the decoder output and used with the first $I'$ additive synthesizers.

\section{Experimental protocol}
\label{sec:method}
In this section, we detail the datasets, metrics and baselines used to assess the performance of the proposed BWE models. 

The task that we consider is bandwidth extension task where the input signal is sampled at $4$kHz, thus with frequencies up to $2$kHz and the output signal is sampled at $16$kHz, thus with frequencies up to $8$kHz. 

As our approach is quite flexible in terms of extension scenario, we also performed experiments for the task going from a sampling frequency of $8$kHz to $16$kHz. We found that the ranking between models was the same as the one for upsampling from 2kHz to 8kHz. We thus display and discuss results only for the latter, as the task is more challenging and lead to more salient perceptual differences, a required aspect for a successful perceptual evaluation.

\subsection{Datasets}

To train and evaluate our models, we used both monophonic and polyphonic datasets. Synthetic data has been also been considered in order to check expected behaviors of proposed systems. Those systems are then evaluated on uncontrolled real-world data.

\subsubsection*{Synthetic datasets}

In order to analyze the inference capabilities of the trained models, we generated two synthetic datasets, respectively containing monophonic and polyphonic signals. These signals are generated using a harmonic-plus-noise synthesizer, as for the DDSP models, allowing for precise analysis of the models generating capabilities.

Each monophonic signal is generated given a $f_0$ corresponding to a certain MIDI pitch between \textit{C3} (\emph{e.g.}, $130.82$~Hz) and \textit{G\#6} (\emph{e.g.}, $1661.22$~Hz). An harmonic signal is generated from this $f_0$ with $H$ harmonics ($H \in \{10, 15, 20\}$), where the amplitude of the $h$-th harmonic is $\frac{1}{h^2}$. A pink noise is added to this harmonic signal with a signal-to-noise ratio of $10$~dB. Then, an \textit{attack, sustain, decay} (ASD) envelope is generated and multiplied to the harmonic-plus-noise signal. The durations of attack and decay and the sustain level are randomly picked in the interval $[0, 0.3]$, $[0.5, 1]$ and $[0, 2]$ (in seconds), respectively. Finally, a random gain in interval $[0.75,1]$ scales the final monophonic harmonic-plus-noise signal. The final monophonic synthetic dataset is obtained by generating all combinations of $f_0$ with the three $H$ values, giving three signals.

The polyphonic synthetic dataset is generated by composing chords on the diatonic scale simply by considering multiple notes from the monophonic synthetic dataset, as follows. To generate a $I$-note polyphonic chord signal, we randomly pick $I$ monophonic signals by taking care that a particular pitch (regardless of the octave) does not appear more than once among these $I$ signals. For each note, a gain is randomly picked in $[0.5, 1]$, and all notes are mixed with corresponding gains. To build the full database, we generated polyphonic signals for all combinations of $f_0$ and $I \in \{2, 3, 4, 5\}$.

From the generated monophonic and polyphonic synthetic datasets, $90$\% of the signals form the train set, and the remaining signals form the test set.

\subsubsection*{Real-world monophonic datasets}

Two real-world datasets consisting of monophonic musical signals are used to evaluate our models. The OrchideaSOL dataset \cite{cella_orchideasol_2020} includes signals of single notes from many different instruments (accordion, bassoon, tuba, horn, trombone, trumpet, guitar, harp, contrabass, viola, violin, violoncello, clarinet, flute, oboe and saxophone). In the original dataset, many different playing styles are available for each instrument, however we only keep the \textit{ordinario} one, corresponding to a natural playing. The training set for our experiments contains $90\%$ of the original dataset, \emph{i.e.}, about $5.5$ hours of audio, while the test set contains $10\%$, \emph{i.e.}, about $42$ minutes of audio.

Medley-solos-db \cite{lostanlen_deep_2017} is another largely monophonic dataset which contains melodies of one of eight different instruments (clarinet, distorted electric guitar, female singer, flute, piano, saxophone, trumpet and violin), \emph{i.e.} the $f_0$ changes over time in those signals. In our experiments we considered the original provided test and train splits, which corresponds to about $2.4$ and $5$ hours of audio, respectively. As some of the instruments are polyphonic \emph{i.e.} distorted electric guitar, piano and violin, a small part of the dataset cannot strictly be considered as monophonic. In order to preserve the integrity of train/test splits of the dataset, we chose not to discard those instruments.

\subsubsection*{Real-world polyphonic datasets}

To assess the proposed model for polyphonic BWE, we employed two real-world datasets containing multiple multi-track mixes. Gtzan dataset \cite{sturm_gtzan_2013} has been widely exploited in many audio signal processing tasks. It contains 1000 30-second music track equally split into 10 genres (blues, classical, country, disco, hip-hop, jazz, metal, pop, reggae and rock). The train and test splits contain $7.5$ hours and $50$ minutes of audio, respectively. 

We also used the mixed version of each track of the MedleyDB dataset \cite{bittner_medleydb_2016}, since most of the corresponding stems are already part of the training split of the previously mentionned Medley-solos-db dataset. The whole MedleyDB dataset is split into train and test sets in a $90\%$/$10\%$ way, corresponding to approximately $6$ hours and $50$ minutes of audio data, respectively.

\subsection{Evaluation and metrics}

To evaluate the performance of the proposed models, we first employ an objective metric computed in the frequency domain named log-spectral distance (LSD), defined as:

\begin{equation}
    LSD = \frac{1}{T} \sum_{t=1}^{T} \sqrt{\frac{1}{K} \sum_{k=1}^{K} (log|Y(n,k)|^2 - log|\tilde{Y}(n,k)|^2)},
\end{equation}

where $Y(n,k)$ and $\tilde{Y}(n,k)$ are the STFT representation of the target full-band signal and the estimated full-band signal, respectively (see Section~\ref{sec:experimental_parameters} for STFT parameter values).

Secondly, we ran a listening test based on a MUSHRA methodology \cite{schoeffler_webmushra_2018} to assess the perceptive accuracy of the proposed models. We followed the classical MUSHRA specifications to build a listening test which was completed by 44 participants. More details are provided in Section \ref{sec:evaluation}.

\subsection{Reference methods}

\subsubsection{Null baseline}

The null baseline is simply the absence of addition of any content in the missing high frequency range. It provides a "ground floor" baseline to assess if the contribution of a given method is not actually worse than doing nothing.

\subsubsection{Spectral band replication}

To compare the performance of our models with existing approaches in the literature, a simplified version of the SBR algorithm \cite{meltzer_sbr_2002} has been evaluated on the considered datasets. This algorithm has a long history in audio codec technologies and comes in various designs that often considers the use of side information, transmitted in the bitstream for the decoder to perform BWE. In this work we implemented a simplified version that is blind, \textit{i.e.} do not require any information for performing BWE.

In this algorithm, the input signal $x$ is treated in the frequency domain, frame by frame. The upper half frequencies are inferred by replicating the lower half frequencies, with the idea of transposing the lower harmonics upwards. As we aim to extend the bandwidth of musical signal from $2$~kHz to $8$~kHz, we extend this algorithm by replicating the lower band three times to reconstruct the full spectrum. In order to obtain a typical frequency amplitude decay, for each replication, the amplitudes of the transposed frequencies are adjusted so that there is an energy continuity at the replication frontier, \emph{i.e.}, for the $j$-th replication ($j \in \{1, 2, 3\}$ as it is a fourfold bandwidth extension), the energies of the same portion of frequencies on both size are equal:

\begin{equation}
    \sum_{k=(j-\alpha) \frac{K}{4}}^{j\frac{K}{4}} |X(k)|^2 = \sum_{k=j\frac{K}{4}}^{(j+\alpha) \frac{K}{4}} |X(k)|^2,
\end{equation}

where $K$ is the number of frequency bins, and $\alpha \in [0, 1]$ the fraction of frequency bins considered for matching the energies of adjacent replicated bands. Experimentally, we found that $\alpha = 0.5$ led to the best performance for the overall algorithm. 

In our experiment, to consider the SBR at its best performance, the ground-truth phase information is used to obtain the full-band signal in the temporal domain. We acknowledge that the phase is not known in practice and would have to be estimated in a realistic production setting.

\subsubsection{Resnet architecture}

We also compare our models to a higher complexity system based on deep learning \cite{sulun_filter_2021}. We chose this system because the Resnet architecture shows better results than the other proposed model based on a U-Net architecture. The Resnet architecture takes an input signal in the temporal domain and output a signal of the same size, with high frequency components. It is composed of 15 residual blocks made of two 1D convolutional layers each, with 512 convolutional filters of size 7, with a rectified linear unit (ReLU) activation after the first layer. For each layer, the input is added back to the output after being multiplied by a factor $0.1$ (for stabilizing the training) in a residual fashion. The input signal is added back to the output. Batch normalization and dropout with a factor $0.5$ are used after each convolutional layers. This model has around 55M learnable parameters.

To train this model, we use the same strategy as in \cite{sulun_filter_2021}, \emph{i.e.}, using a mean square error loss with a learning rate reducing schedule.

\subsection{Experimental parameters}
\label{sec:experimental_parameters}
In our work, all audio signals are sampled at $f_s=16$~kHz. To compute the STFT of these signals, we used an analysis window of $1024$ samples with a hop length of $256$ samples. The input signals are of length $64000$ samples ($=4$~s) for the DDSP models and the SBR baseline, and $8192$ samples ($\approx 0.5$~s) for the Resnet model. In the DDSP models, we considered $H=100$ number of harmonics and a size of $K=65$ for the noise transfer function $N(k)$. In the cyclic DDSP system, we use a total number of $I=5$ iterations.

DDSP models are trained for $25000$ steps with batches of size $32$. We used the Adam optimizer with an initial learning rate of $0.001$ for DDSP models, and the latter is halved if the loss has not been decreased during four plateaus of $2500$ steps. We used A100 GPUs for the training, which permit us to train DDSP models for around $1$ hour for \textit{DDSP-mono-dec}, $2$ hours for \textit{DDSP-poly-dec}, while Resnet training took around $19$ hours.

\begin{figure*}[ht!]
\centering
\includegraphics[width=1.\columnwidth]{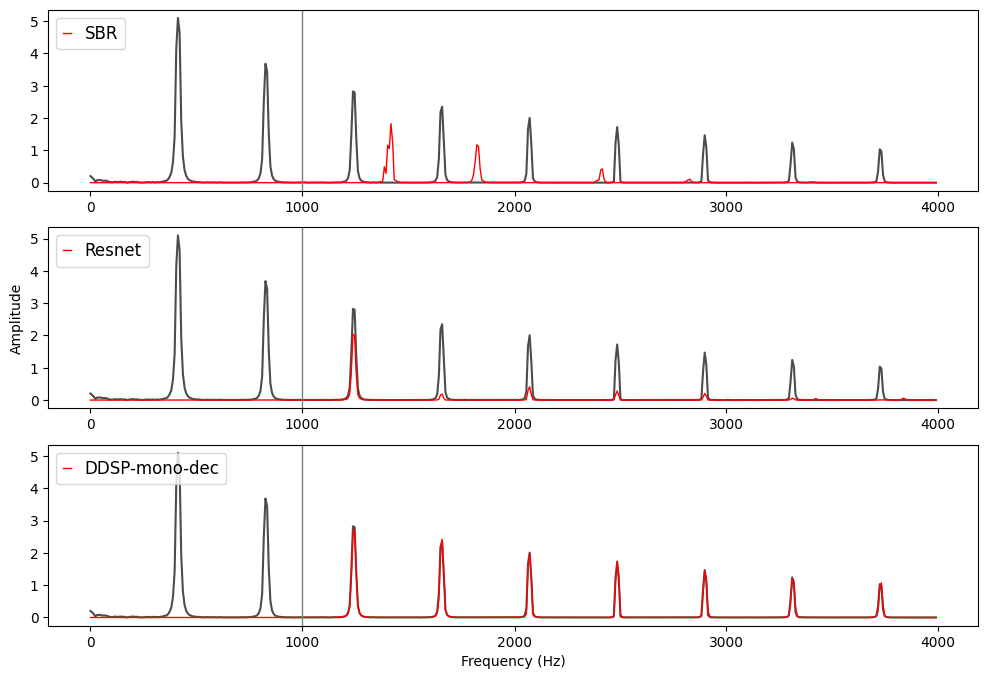}
\caption{Generated upper frequency band using the model \textit{DDSP-mono-dec}, the baseline SBR and the reference Resnet model, for a synthetic signal containing harmonics based on the MIDI note G\#5 ($\approx 830$~Hz). The vertical line shows the limit between the low and high bands.}
\label{fig:spectrum_mono_synth}
\end{figure*}

\section{Validation on synthetic data}
\label{sec:synthetic_data}

In this section, we first study the performance of the proposed models against the baselines on the monophonic and polyphonic synthetic data. It allows for more detailed insights on the models' ability to accurately generate the missing high frequency content.

\subsection{Monophonic dataset}

We first trained and evaluated our monophonic DDSP model on the monophonic synthetic dataset against SBR method \cite{meltzer_sbr_2002} and Resnet model \cite{sulun_filter_2021}. Table \ref{tab:synth_mono} summarises the results.

\begin{table}[ht]
\centering
\caption{Evaluation results for monophonic BWE model and baselines on the monophonic synthetic dataset.}
\begin{tabular}{l|c}
\textbf{Model}      & \textbf{LSD} \\ \hline
\textbf{\textbf{Null}}        &  6.15        \\
\textbf{SBR \cite{meltzer_sbr_2002}}        & 4.19         \\
\textbf{Resnet \cite{sulun_filter_2021}}     & 4.34         \\
\textbf{DDSP-noise}     & 5.04         \\
\textbf{DDSP-mono-dec}  &\textbf{2.93}         \\ 
\end{tabular}

\label{tab:synth_mono}
\end{table}

The results show the benefit of the DDSP model over the two reference models. On Fig.~\ref{fig:spectrum_mono_synth}, the generated upper band from the proposed monophonic DDSP model, SBR baseline and Resnet models are illustrated for one frame of a particular synthetic signal with $f_0 \approx 830$~Hz. The DDSP model is robust enough to synthesize the wanted harmonics with matching amplitudes, showing that it is capable to learn the chosen harmonic amplitude decay. The SBR baseline duplicates the low-band harmonic content with an offset because of the mismatch between the cutoff frequency and $f_0$, and the Resnet model is apparently not capable of generating relevant high frequency harmonics, thus minimizing its loss by very few addition of energy.

\subsection{Polyphonic dataset}

\begin{figure*}[ht!]
\centering
\includegraphics[width=1.\columnwidth]{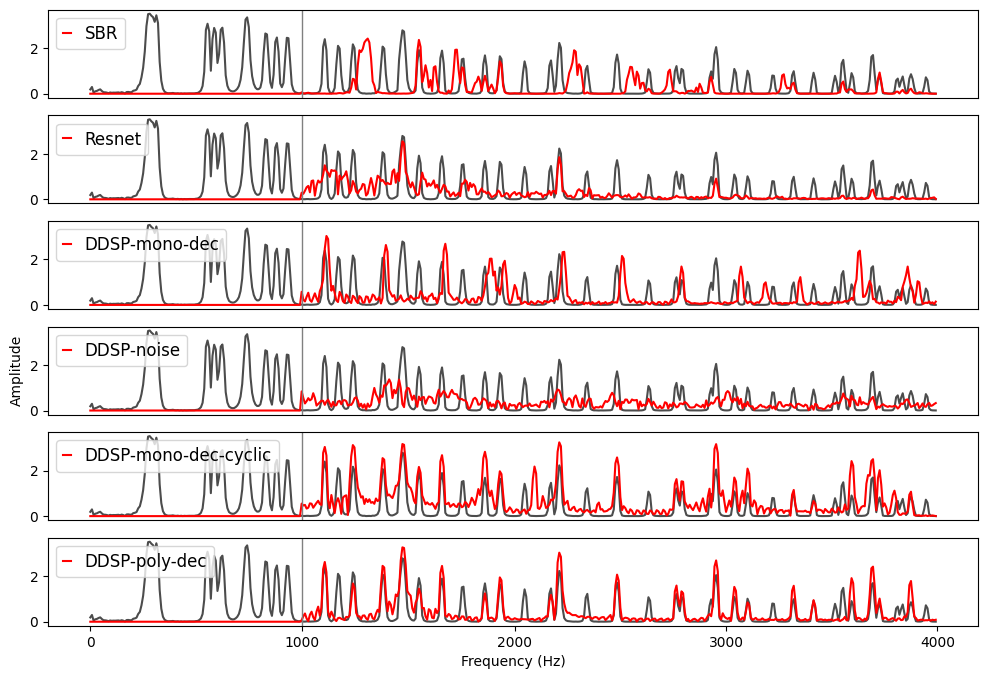}
\caption{Generated upper frequency band using the proposed models and the baselines, for one frame of a synthetic signal containing four notes: C\#5 ($\approx 554$~Hz), D\#5 ($\approx 587$~Hz), D5 ($\approx 622$~Hz) and F\#6 ($\approx 1479$~Hz). The vertical line shows the limit between the low and high bands.}
\label{fig:spectrum_poly_synth}
\end{figure*}

\begin{table}[ht]
\centering
\caption{Evaluation results for the proposed BWE models, the baseline and the reference resnet model on the polyphonic synthetic datasets.}
\begin{tabular}{l|c}
\textbf{Model}       & \textbf{LSD} \\ \hline
\textbf{Null}         &  11.03       \\
\textbf{SBR \cite{meltzer_sbr_2002}}         & 8.77         \\
\textbf{Resnet \cite{sulun_filter_2021}}      & 8.37         \\
\textbf{DDSP-mono-dec}   & 7.86         \\
\textbf{DDSP-noise}  & 9.18             \\
\textbf{DDSP-mono-dec-cyclic} & 5.59         \\
\textbf{DDSP-poly-dec}   & \textbf{4.72}         \\ 
\end{tabular}
\label{tab:synth_poly}
\end{table}

Table~\ref{tab:synth_poly} shows the LSD metric for all models and baselines on the polyphonic synthetic dataset. We can see that all proposed models surpass the SBR baseline and the Resnet model, and that the BWE performance has been improved by the design of both the cyclic system and the polyphonic model. The polyphonic DDSP model is almost twice as good as SBR and Resnet, which is an important improvement. When looking at Fig.~\ref{fig:spectrum_poly_synth}, which illustrated the upper band generation for one polyphonic example from all the considered models and baselines, we notice that both cyclic and polyphonic methods are capable of generating precise harmonics, with a relatively good amplitude match compare to the ground-truth. For the monophonic setting, the SBR baseline generates shifted harmonics. The Resnet model seems to be able to focus only on some harmonics, with relatively precise amplitudes, while also generating some noise in the lowest generated frequencies. 

The three DDSP-based models seems quite capable of estimating the low harmonic amplitudes, while the high harmonic content suffers from too high amplitudes, which may lead to non-natural artifacts. Possible reasons for this defect are given at the end of the next section. 

\section{Evaluation on real-world datasets}
\label{sec:evaluation}

In this section, we present the performance results for each monophonic and polyphonic recorded datasets of the proposed models against the reference methods: namely SBR and Resnet model.

\subsection{Objective evaluation}

The proposed models are first evaluated objectively using the LSD metrics on the real-world datasets. Monophonic models \textit{DDSP-mono-dec} and \textit{DDSP-noise} are evaluated on both monophonic and polyphonic datasets, while polyphonic models \textit{DDSP-mono-dec-cyclic} and \textit{DDSP-poly-dec} are evaluated only on polyphonic datasets (\textit{Gtzan} and \textit{MedleyDB}. Table~\ref{tab:all_datasets} shows the results.

\begin{table*}[ht]
\centering
\caption{LSD performance of the evaluated models for monophonic and polyphonic real-world datasets. Best models are shown in bold for each dataset. The last two columns show CPU inference time expressed as real time percentage and the number of parameters of the models.}
\begin{adjustbox}{width=1.\textwidth}
\begin{tabular}{l|cccc|c|c}
\multicolumn{1}{c|}{\multirow{3}{*}{\textbf{Model}}} & \multicolumn{4}{c|}{\textbf{Log-spectral distance}}                                                                                                                    & \multicolumn{1}{c|}{\multirow{2}{*}{\textbf{Inference time}}} & \multicolumn{1}{c}{\multirow{3}{*}{\textbf{\# of parameters}}} \\ \cline{2-5}
\multicolumn{1}{c|}{}                                & \multicolumn{2}{c|}{\textbf{Monophonic datasets}}                                        & \multicolumn{2}{c|}{\textbf{Polyphonic datasets}}                           & \multicolumn{1}{c|}{} & \multicolumn{1}{c}{}                                                       \\ \cline{2-5}
\multicolumn{1}{c|}{}                                & \multicolumn{1}{l}{\textbf{OrchideaSOL}} & \multicolumn{1}{l|}{\textbf{Medley-solos-db}} & \multicolumn{1}{l}{\textbf{MedleyDB}} & \multicolumn{1}{l|}{\textbf{Gtzan}} & \multicolumn{1}{c|}{\textbf{(\% real-time)}} & \multicolumn{1}{c}{}                                                       \\ \hline
\textbf{Null}                                         & 15.9                                      & \multicolumn{1}{c|}{18.53}                     &  24.37                               & 33.84                               & 0 & 0                                                                           \\
\textbf{SBR} \cite{meltzer_sbr_2002}                                         & 9.27                                     & \multicolumn{1}{c|}{8.78}                     & 11.15                                 & 12.96                               & 2  & 0                                                                          \\
\textbf{Resnet} \cite{sulun_filter_2021}                                      & 14.04                                    & \multicolumn{1}{c|}{15.65}                    & 16.17                                 & 26.84                               & 48 & 55M                                                                        \\ \hline
\textbf{DDSP-noise}                                   & {7.20}                            & \multicolumn{1}{c|}{{8.28}}            & \textbf{8.96}                         & 10.06                               & 3 & 3.5M                                                                           \\
\textbf{DDSP-mono-dec}                                & \textbf{5.68}                            & \multicolumn{1}{c|}{\textbf{8.09}}            & 8.98                                  & \textbf{9.95}                       & 9    & 4.4M                                                                        \\ \hline
\textbf{DDSP-mono-dec-cyclic}                         & /                                        & \multicolumn{1}{c|}{/}                        & 11.57                                 & 11.60                               & 44 & 4.4M                                                                          \\
\textbf{DDSP-poly-dec}                                & /                                        & \multicolumn{1}{c|}{/}                        & 9.53                                  & 10.31                               & 9   & 7.5M                                                                         \\ 
\end{tabular}
\end{adjustbox}
\label{tab:all_datasets}
\end{table*}

First, we can see that all proposed models surpass both SBR and Resnet model in terms of LSD, except for the cyclic model which is worse than SBR. On the OrchideaSOL, Gtzan and MedleyDB datasets, the gain in performance is substantial for the best model compared to the reference ones. For example, \textit{DDSP-mono-dec} leads to a LSD of $5.68$ where SBR and Resnet achieve $9.27$ and $14.04$, respectively. On polyphonic signals, the Resnet model seems to be quite bad at predicting high frequencies (LSD = $26.84$ and $16.17$ on Gtzan and MedleyDB, respectively), whereas our DDSP-based models give quite lower LSDs (less than $12$ for all theses models on both datasets).

\begin{figure*}
\centering
\includegraphics[width=1.\columnwidth]{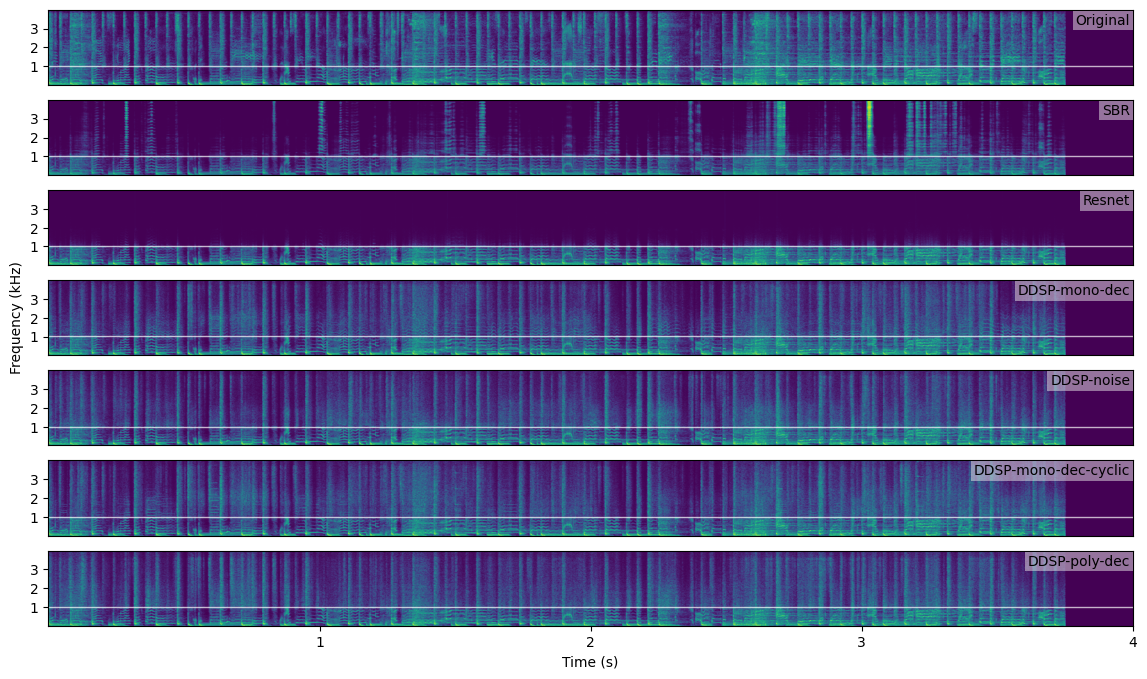}
\caption{Spectrograms showing the generated upper frequency band using the proposed models and the baselines for a real-world signal from a pop music track. The horizontal white line shows the limit between the low and high bands at $1000$ Hz.}
\label{fig:spectrum_poly_real}
\end{figure*}

When looking at the performance of the proposed models, we first observe that the polyphonic models \textit{DDSP-mono-dec-cyclic} and \textit{DDSP-poly-dec} do not achieve a better performance than the monophonic one \textit{DDSP-mono-dec}, whereas the noise-only model \textit{DDSP-noise} is on par with its results. This observation is quite constrasted from what we obtained on the synthetic datasets. When having a look at Fig.~\ref{fig:spectrum_poly_real}, we can see that \textit{DDSP-poly-dec} seems to generate the highest frequency with too low amplitudes, whereas \textit{DDSP-mono-dec} is a bit more precise in the high frequencies. 

By informal listening of some reconstructed signals, we managed to distinguish two types of unwanted artifacts. The first kind happens when the amplitudes of the reconstructed harmonics are too high, which leads to a very synthetic high frequency reconstruction. One of the reasons for these wrongly inferred harmonic amplitudes is that, in both \textit{DDSP-mono-dec-cyclic} and \textit{DDSP-mono-dec}, the loudness contour is estimated for a mixture made of several $f_0$, making it less trivial for the autoencoder to estimate each $f_0$ harmonic amplitudes. The second type of artifacts can be heard when the synthesized noise handles much of the high frequency content, while the harmonic amplitudes are too low, or even non-existent. This happens when the multi-pitch estimator fails to correctly predict the set of $f_0$s, then the overall system do not generate high amplitude harmonics, and compensates with noise. Because of that, we conjecture that the proposed models should be more effective with a more robust multi-pitch estimation system.

\subsection{Perceptual evaluation}

\begin{figure}[ht!]
\centering
\includegraphics[width=.7\columnwidth]{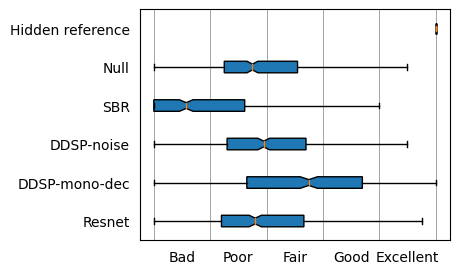}
\caption{Stimuli ratings for, from top to bottom, the anchor \textit{Null} corresponding to the input signal without any process, the \textit{SBR} baseline, the proposed models \textit{DDSP-mono-dec} and \textit{DDSP-noise}, and the reference \textit{Resnet} model. Boxes correspond to the interquartile range (IQR) over all participants, with the mean indicated by an orange vertical line. Lower and upper whiskers are set to $1.5 \times IQR$ below and above $Q_1$ and $Q_3$, respectively.}
\label{fig:boxplot_mushra}
\end{figure}

In order to assess the perceptive value of our models, we conducted a listening test based on the MUSHRA methodology \cite{schoeffler_webmushra_2018}. During the listening test, 42 subjects were asked to rate the quality of audio signals between $0$ (poor quality) and $100$ (perfect quality) against the reference (ground-truth full-band signal), which is expected to be rated $100$. This behavior is expected by normal hearing and focused subjects, as the reference sound is provided for each trial. $10$ stimuli are given in a random order. For each of them, $6$ signals are to be rated :
\begin{enumerate}
    \item Anchor 1: low-band input signal (model \textit{Null})
    \item Anchor 2: hidden reference (ground-truth full-band signal)
    \item SBR reconstruction
    \item Resnet output
    \item DDSP-mono-dec output
    \item DDSP-noise output
\end{enumerate}

The signals are taken from Gtzan dataset \cite{sturm_gtzan_2013}, one of each genre, and only 5 seconds are extracted in the middle of the original signal. Information about participants are asked at the end of the survey, including gender, age, and the number of years of musical practice. Given the poorer LSD performance of our proposed polyphonic schemes compared to the monophonic one, also confirmed by informal listening by the authors, it has been decided not to consider them for subjective evaluation. This had the benefit of maintaining the duration of the listening test into a reasonable range of about 20 minutes.

First, we conduct an analysis of variance (ANOVA) to check whether the factor of musical training is a significant source of variation in the rating data. We consider an individual as being a musician if it has an experience of at least one year. Considering that, the ANOVA run on the rating distributions for musician and non-musician subjects gives a $p$-value of $1.61 \cdot 10^{-8}$, which tells us that being a musician or not has a significant effect on the test ratings. 

\begin{figure}[ht!]
\centering
\includegraphics[width=.6\columnwidth]{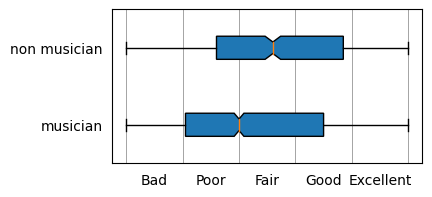}
\caption{Stimuli ratings from subjects with (bottom) and without (top) musical training. Boxes correspond to the interquartile range (IQR) over all participants, with the mean indicated by an orange vertical line. Lower and upper whiskers are set to $1.5 \times IQR$ below and above $Q_1$ and $Q_3$, respectively.}
\label{fig:musicians}
\end{figure}

Close inspection of the ratings showed that the rankings of the different methods are the same for both populations. The only difference a different bias, where musicians were on average more severe than non-musicians, as can be seen on Fig.~\ref{fig:musicians}, that show the distributions of ratings for musicians and non-musicians the over all models.

Next, conducting another ANOVA in which the analyzed factor is the model gives us a $p$-value of $2.18.10^{-88}$, which is very small and shows that the choice of model is a significant source of variation in our collected data, thus the possibility of comparing the rating distributions of all models. With another ANOVAs on the models but for the data splits in musician or non-musician subsets, we obtain similar very low-valued $p$-values, which tells us that in both case the choice of models has an significant impact on the ratings among the participants.

The distributions of the participants' ratings for all models and all stimuli are plotted as boxplots in Fig.~\ref{fig:boxplot_mushra}. We can see that the outputs of model \textit{DDSP-mono-dec} are in average rated to be of \textit{fair} quality (almost \textit{good}), whereas the outputs from \textit{Resnet}, \textit{DDSP-noise} and no processing are typically rated as \textit{poor}, and \textit{SBR} outputs are quite often rated as being of \textit{bad} quality. An important outcome is that \textit{DDSP-mono-dec} provides a large margin improvement compared to the \textit{Null} baseline, meaning that this method is able to improve audio quality at a low computational cost.

By computing a \textit{t}-test on rating distribution of \textit{DDSP-mono-dec} against the other models, we can verify that it is significantly better than the other ones. The $p$-values obtained for the \textit{t}-test are well below the typical threshold of $0.05$, so the distributions are significantly different from each other. We can thus conclude that, from the data of the listening test, the \textit{DDSP-mono-dec} gives perceptively better high-frequency contents than the other evaluated models.

Monophonic and polyphonic examples are available online\footnote{\url{https://mathieulagrange.github.io/ddspMusicBandwidthExtension}}. The latter have been considered as stimuli in the listening test.

\subsection{Inference time}

\begin{figure}[ht!]
\centering
\includegraphics[width=0.5\columnwidth]{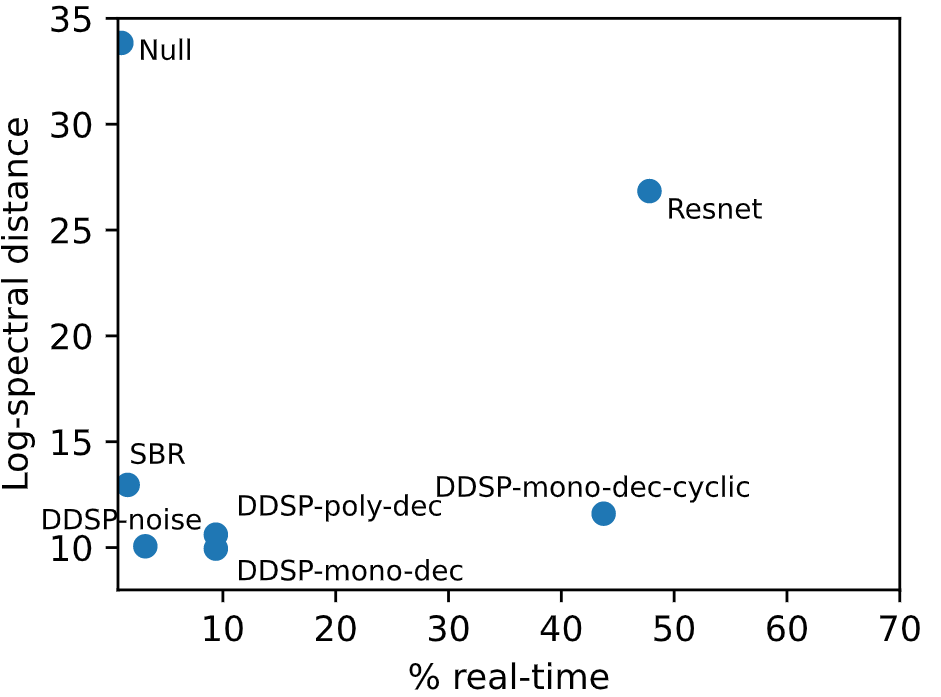}
\caption{Real-time CPU inference percentage vs. LSDs for the proposed models and the baselines over the dataset Gtzan.}
\label{fig:lsd_vs_time}
\end{figure}

One great advantage of the proposed DDSP approach is the important reduction of inference time compared to neural networks with a lot of training parameters such as Resnet. Fig.~\ref{fig:lsd_vs_time} shows a scatter plot of the performance against the inference time of the different proposed models and the baselines SBR and Resnet, on the Gtzan dataset. The inference were made on a laptop equipped with a Intel Core i7 CPU at frequency of $2.8$ GHz.  We can clearly see that a neural network architecture such as Resnet takes a lot of computing time to process an input signal, well above a potential real-time behavior. While SBR is very fast, DDSP-based models such as \textit{DDSP-mono-dec}, \textit{DDSP-poly-dec} and \textit{DDSP-noise} are quite efficient in terms of computation time. \textit{DDSP-mono-dec} and \textit{DDSP-poly-dec} takes the same amount of computing time because their architecture is very similar, and \textit{DDSP-noise} is a bit faster because of smaller matrices in the decoder. On the other hand, \textit{DDSP-mono-dec-cyclic} is less computationally efficient because of its iterative nature, as an inference from \textit{DDSP-mono-dec} is computed at each iteration. These insights on the computational power of DDSP-based model show the advantage of such hybrid models compared to neural networks with a huge number of parameters such as the Resnet architecture.

\section{Conclusion}
\label{sec:conclusion}
In this article, we explored differentiable digital signal processing models for bandwidth extension of monophonic and polyphonic musical signals. We showed the benefit of using a monophonic DDSP model to generate high frequencies of monophonic signals against the two baselines, including a high complexity deep-learning-based resnet model. Then, we designed two systems to address polyphonic BWE: a cyclic use of a monophonic DDSP model, and an adapted DDSP model with polyphonic synthesis capacities. On polyphonic signals, the proposed polyphonic systems showed to be more effective on polyphonic synthetic signals, but failed to surpass the monophonic DDSP model on real data. In addition, we conducted a listening test with the MUSHRA methodology, which showed that the \textit{DDSP-mono-dec} model was more pleasant to the ear for most participants, when compared to the baselines. For future work, we think that considering a more advanced multi-pitch estimator could enable the polyphonic models to generate less artifacts, and that other artifacts could be avoided by researching further the loudness estimation procedure.

\section*{Availability of data and materials}

Experiments reported in this paper rely on publicly available code and on the following publicly available datasets:
\begin{enumerate}
    \item \textbf{OrchideaSOL}: \url{https://forum.ircam.fr/projects/detail/orchideasol}
    \item \textbf{Medley-solos-db}: \url{https://zenodo.org/record/1344103}
    \item \textbf{MedleyDB}: \url{https://medleydb.weebly.com}
    \item \textbf{GTZAN}: \url{https://www.kaggle.com/datasets/andradaolteanu/gtzan-dataset-music-genre-classification}
\end{enumerate}
The synthetic datasets can be reproduced using the experimental code available at: \url{https://github.com/mathieulagrange/ddspMusicBandwidthExtension}.

\section*{Acronyms}
\begin{table}[h!]
\centering
\caption{Table of acronyms.}
\label{tab:acronyms}
    \begin{tabular}{|lr|}
    \hline
        \textbf{ANOVA}         & analysis of variance \\
        \textbf{ASD}         & attack, sustain, release \\
        \textbf{ASR}         & automatic speech recognition \\
        \textbf{BWE}      & bandwidth extension \\
        \textbf{CNN}         & convolutional neural network \\
        \textbf{CPU}         & central processing unit \\
        \textbf{DDSP}     & differentiable digital signal processing \\
        \textbf{DL}         & deep learning \\
        \textbf{DNN}         &  deep neural network \\
        \textbf{FFT}         & fast Fourier transform \\
        \textbf{FIR}         & finite impulse response \\
        \textbf{GAN}         & generative adversarial network \\
        \textbf{GBRBM}         &  Gaussian-Bernoulli restricted Boltzmann machine \\
        \textbf{GMM}         & Gaussian mixture model \\
        \textbf{GPU}         & graphics processing unit \\
        \textbf{GRU}         & gated recurrent unit \\
        \textbf{HMM}         & hidden Markov model \\
        \textbf{LB}         & low-band \\
        \textbf{LSD}         & log-spectral distance \\
        \textbf{LSTM}         & long short-term memory \\
        \textbf{MFCC}         & mel frequency cepstral coefficients \\
        \textbf{MLP}         & multi-layer perceptron \\
        \textbf{MSE}         & mean-squared error \\
        \textbf{MSS}         & multi-scale spectral \\
        \textbf{MUSHRA}         & multiple stimuli with hidden reference and anchor \\
        \textbf{NMF}         & non-negative matrix factorization \\
        \textbf{ReLU}         & rectified linear unit \\
        \textbf{SBR}      & spectral band replication  \\
        \textbf{STFT}         & short-term Fourier transform \\
        \textbf{WB}         & wide-band \\
    
    \hline
    \end{tabular}
\end{table}

\section*{Competing interests}
  The authors declare that they have no competing interests.

\section*{Funding}
  This research have been partially funded by an RFI OIC grant. 

\section*{Author's contributions}
    PAG conducted the numerical experiments and wrote the manuscript. ML provided guidance and wrote the manuscript.

\section*{Acknowledgements}
  We would like to thank Vincent Lostanlen for fruitful discussions and suggestions.

\bibliographystyle{apalike}
\bibliography{biblio}  

\begin{thebibliography}{}

\bibitem[Adler et~al., 2012]{adler_audio_2012}
Adler, A., Emiya, V., Jafari, M.~G., Elad, M., Gribonval, R., and Plumbley, M.~D. (2012).
\newblock Audio inpainting.
\newblock {\em IEEE Transactions on Audio, Speech, and Language Processing}.

\bibitem[Bansal et~al., 2005]{bansal_bandwidth_2005}
Bansal, D., Raj, B., and Smaragdis, P. (2005).
\newblock Bandwidth expansion of narrowband speech using non-negative matrix factorization.
\newblock In {\em Interspeech}.

\bibitem[Bauer and Fingscheidt, 2008]{bauer_hmm-based_2008}
Bauer, P. and Fingscheidt, T. (2008).
\newblock An {HMM}-based artificial bandwidth extension evaluated by cross-language training and test.
\newblock In {\em International {Conference} on {Acoustics}, {Speech} and {Signal} {Processing}}.

\bibitem[Bittner et~al., 2022]{bittner_lightweight_2022}
Bittner, R.~M., Bosch, J.~J., Rubinstein, D., Meseguer-Brocal, G., and Ewert, S. (2022).
\newblock A lightweight instrument-agnostic {Model} for polyphonic note transcription and multipitch estimation.
\newblock In {\em International {Conference} on {Acoustics}, {Speech} and {Signal} {Processing}}.

\bibitem[Bittner et~al., 2016]{bittner_medleydb_2016}
Bittner, R.~M., Wilkins, J., Yip, H., and Bello, J.~P. (2016).
\newblock {MedleyDB} 2.0: {New} data and a system for sustainable data collection.
\newblock In {\em International Conference on Music Information Retrieval}.

\bibitem[Campos et~al., 2018]{campos_high_2018}
Campos, G., Fonseca, N., Ferreira, A., and Davies, M. (2018).
\newblock High frequency magnitude spectrogram reconstruction for music mixtures using convolutional autoencoders.
\newblock In {\em Conference on {Digital} {Audio} {Effects}}.

\bibitem[Cella et~al., 2020]{cella_orchideasol_2020}
Cella, C.~E., Ghisi, D., Lostanlen, V., Lévy, F., Fineberg, J., and Maresz, Y. (2020).
\newblock {OrchideaSOL}: a dataset of extended instrumental techniques for computer-aided orchestration.

\bibitem[Chennoukh et~al., 2001]{chennoukh_speech_2001}
Chennoukh, S., Gerrits, A., Miet, G., and Sluijter, R. (2001).
\newblock Speech enhancement via frequency bandwidth extension using line spectral frequencies.
\newblock In {\em International {Conference} on {Acoustics}, {Speech}, and {Signal} {Processing}. {Proceedings}}.

\bibitem[Dietz et~al., 2002]{dietz_spectral_2002}
Dietz, M., Liljeryd, L., Kjorling, K., and Kunz, O. (2002).
\newblock Spectral band replication, a novel approach in audio coding.
\newblock In {\em Audio {Engineering} {Society} {Convention}}.

\bibitem[Engel et~al., 2020]{engel_ddsp_2020}
Engel, J., Hantrakul, L., Gu, C., and Roberts, A. (2020).
\newblock {DDSP}: {Differentiable} {Digital} {Signal} {Processing}.

\bibitem[Esqueda et~al., 2021]{esqueda_differentiable_2021}
Esqueda, F., Kuznetsov, B., and Parker, J.~D. (2021).
\newblock Differentiable white-box virtual analog modeling.
\newblock In {\em International {Conference} on {Digital} {Audio} {Effects}}.

\bibitem[French and Steinberg, 1947]{french_factors_1947}
French, N.~R. and Steinberg, J.~C. (1947).
\newblock Factors governing the intelligibility of speech sounds.
\newblock {\em The Journal of the Acoustical Society of America}.

\bibitem[Gaultier et~al., 2021]{gaultier_sparsity-based_2021}
Gaultier, C., Kitić, S., Gribonval, R., and Bertin, N. (2021).
\newblock Sparsity-based audio declipping methods: selected overview, new algorithms, and large-scale evaluation.
\newblock {\em IEEE/ACM Transactions on Audio, Speech, and Language Processing}.

\bibitem[Gu and Ling, 2017]{gu_waveform_2017}
Gu, Y. and Ling, Z.-H. (2017).
\newblock Waveform modeling using stacked dilated convolutional neural networks for speech bandwidth extension.
\newblock In {\em Interspeech}.

\bibitem[Hantrakul et~al., 2019]{hantrakul_fast_2019}
Hantrakul, L., Engel, J., Roberts, A., and Gu, C. (2019).
\newblock Fast and flexible neural audio synthesis.
\newblock In {\em International {Society} for {Music} {Information} {Retrieval} {Conference}}.

\bibitem[Hayes et~al., 2021]{hayes_neural_2021}
Hayes, B., Saitis, C., and Fazekas, G. (2021).
\newblock Neural waveshaping synthesis.

\bibitem[Kim et~al., 2018]{kim_crepe_2018}
Kim, J.~W., Salamon, J., Li, P., and Bello, J.~P. (2018).
\newblock Crepe: {A} {Convolutional} {Representation} for {Pitch} {Estimation}.
\newblock In {\em International {Conference} on {Acoustics}, {Speech} and {Signal} {Processing}}.

\bibitem[Kuleshov et~al., 2017]{kuleshov_audio_2017}
Kuleshov, V., Enam, S.~Z., and Ermon, S. (2017).
\newblock Audio super resolution using neural networks.

\bibitem[Lagrange and Gontier, 2020]{lagrange_bandwidth_2020}
Lagrange, M. and Gontier, F. (2020).
\newblock Bandwidth extension of musical audio signals with no side information using dilated convolutional neural networks.
\newblock In {\em International {Conference} on {Acoustics}, {Speech} and {Signal} {Processing}}.

\bibitem[Lee et~al., 2022]{lee_differentiable_2022}
Lee, S., Choi, H.-S., and Lee, K. (2022).
\newblock Differentiable artificial reverberation.
\newblock {\em IEEE/ACM Transactions on Audio, Speech, and Language Processing}.

\bibitem[Li et~al., 2015]{li_dnn-based_2015}
Li, K., Huang, Z., Xu, Y., and Lee, C.-H. (2015).
\newblock {DNN}-based speech bandwidth expansion and its application to adding high-frequency missing features for automatic speech recognition of narrowband speech.
\newblock In {\em Interspeech}.

\bibitem[Li and Lee, 2015]{li_deep_2015}
Li, K. and Lee, C.-H. (2015).
\newblock A deep neural network approach to speech bandwidth expansion.
\newblock In {\em International {Conference} on {Acoustics}, {Speech} and {Signal} {Processing}}.

\bibitem[Li et~al., 2018]{li_speech_2018}
Li, S., Villette, S., Ramadas, P., and Sinder, D.~J. (2018).
\newblock Speech bandwidth extension using generative adversarial networks.
\newblock In {\em International {Conference} on {Acoustics}, {Speech} and {Signal} {Processing}}.

\bibitem[Li et~al., 2019]{li_speech_2019}
Li, X., Chebiyyam, V., and Kirchhoff, K. (2019).
\newblock Speech audio super-resolution for speech recognition.
\newblock In {\em Interspeech}.

\bibitem[Lostanlen and Cella, 2017]{lostanlen_deep_2017}
Lostanlen, V. and Cella, C.-E. (2017).
\newblock Deep convolutional networks on the pitch spiral for musical instrument recognition.

\bibitem[Masuda and Saito, 2021]{masuda_synthesizer_2021}
Masuda, N. and Saito, D. (2021).
\newblock Synthesizer sound matching with differentiable {DSP}.
\newblock In {\em International {Society} for {Music} {Information} {Retrieval} {Conference}}.

\bibitem[Meltzer et~al., 2002]{meltzer_sbr_2002}
Meltzer, S., Bohm, R., and Henn, F. (2002).
\newblock {SBR} enhanced audio codecs for digital broadcasting such as "{Digital} {Radio} {Mondiale}" ({DRM}).
\newblock In {\em Audio {Engineering} {Society} {Convention}}.

\bibitem[Mohammadi and Kain, 2017]{mohammadi_overview_2017}
Mohammadi, S.~H. and Kain, A. (2017).
\newblock An overview of voice conversion systems.
\newblock In {\em Speech {Communication}}.

\bibitem[Moliner et~al., 2023a]{moliner2023zero}
Moliner, E., Elvander, F., and V{\"a}lim{\"a}ki, V. (2023a).
\newblock Zero-shot blind audio bandwidth extension.
\newblock {\em arXiv preprint arXiv:2306.01433}.

\bibitem[Moliner et~al., 2023b]{moliner_solving_2023}
Moliner, E., Lehtinen, J., and Välimäki, V. (2023b).
\newblock Solving audio inverse problems with a diffusion model.
\newblock In {\em International {Conference} on {Acoustics}, {Speech} and {Signal} {Processing}}.

\bibitem[Moliner and Välimäki, 2022]{moliner_behm-gan_2022}
Moliner, E. and Välimäki, V. (2022).
\newblock {BEHM}-{GAN}: {Bandwidth} {Extension} of {Historical} {Music} using {Generative} {Adversarial} {Networks}.

\bibitem[Nagel and Disch, 2009]{nagel_harmonic_2009}
Nagel, F. and Disch, S. (2009).
\newblock A harmonic bandwidth extension method for audio codecs.
\newblock In {\em International {Conference} on {Acoustics}, {Speech} and {Signal} {Processing}}.

\bibitem[Ning et~al., 2019]{ning_review_2019}
Ning, Y., He, S., Wu, Z., Xing, C., and Zhang, L.-J. (2019).
\newblock A review of deep learning based speech synthesis.
\newblock {\em Applied Sciences}.

\bibitem[Oord et~al., 2016]{oord2016wavenet}
Oord, A. v.~d., Dieleman, S., Zen, H., Simonyan, K., Vinyals, O., Graves, A., Kalchbrenner, N., Senior, A., and Kavukcuoglu, K. (2016).
\newblock Wavenet: A generative model for raw audio.
\newblock {\em Proceedings of ISCA}.

\bibitem[Park and Kim, 2000]{park_narrowband_2000}
Park, K.-Y. and Kim, H.~S. (2000).
\newblock Narrowband to wideband conversion of speech using {GMM} based transformation.
\newblock In {\em International {Conference} on {Acoustics}, {Speech}, and {Signal} {Processing}. {Proceedings}}.

\bibitem[Sadasivan et~al., 2016]{sadasivan_joint_2016}
Sadasivan, J., Mukherjee, S., and Seelamantula, C.~S. (2016).
\newblock Joint dictionary training for bandwidth extension of speech signals.
\newblock In {\em International {Conference} on {Acoustics}, {Speech} and {Signal} {Processing}}.

\bibitem[Schoeffler et~al., 2018]{schoeffler_webmushra_2018}
Schoeffler, M., Bartoschek, S., Stöter, F.-R., Roess, M., Westphal, S., Edler, B., and Herre, J. (2018).
\newblock {webMUSHRA} — {A} {Comprehensive} {Framework} for {Web}-based {Listening} {Tests}.
\newblock {\em Journal of Open Research Software}.

\bibitem[Shan et~al., 2021]{shan_differentiable_2021}
Shan, S., Hantrakul, L., Chen, J., Avent, M., and Trevelyan, D. (2021).
\newblock Differentiable wavetable synthesis.
\newblock In {\em International {Conference} on {Acoustics}, {Speech} and {Signal} {Processing}}.

\bibitem[Song and Martynovich, 2009]{song_study_2009}
Song, G.-B. and Martynovich, P. (2009).
\newblock A study of {HMM}-based bandwidth extension of speech signals.
\newblock {\em Signal Processing}.

\bibitem[Song et~al., 2020]{song2020denoising}
Song, J., Meng, C., and Ermon, S. (2020).
\newblock Denoising diffusion implicit models.
\newblock {\em International Conference on Learning Representations}.

\bibitem[Steinmetz et~al., 2022]{steinmetz_style_2022}
Steinmetz, C.~J., Bryan, N.~J., and Reiss, J.~D. (2022).
\newblock Style transfer of audio effects with differentiable signal processing.

\bibitem[Sturm, 2013]{sturm_gtzan_2013}
Sturm, B.~L. (2013).
\newblock The {GTZAN} dataset: {Its} contents, its faults, their effects on evaluation, and its future use.

\bibitem[Su et~al., 2021]{su_bandwidth_2021}
Su, J., Wang, Y., Finkelstein, A., and Jin, Z. (2021).
\newblock Bandwidth extension is all you need.
\newblock In {\em International {Conference} on {Acoustics}, {Speech} and {Signal} {Processing}}.

\bibitem[Sulun and Davies, 2021]{sulun_filter_2021}
Sulun, S. and Davies, M. E.~P. (2021).
\newblock On filter generalization for music bandwidth extension using deep neural networks.
\newblock {\em Journal of Selected Topics in Signal Processing}.

\bibitem[Sun and Mazumder, 2013]{sun_non-negative_2013}
Sun, D.~L. and Mazumder, R. (2013).
\newblock Non-negative matrix completion for bandwidth extension: {A} convex optimization approach.
\newblock In {\em International {Workshop} on {Machine} {Learning} for {Signal} {Processing}}.

\bibitem[Vaseghi and Frayling-Cork, 1992]{vaseghi_restoration_1992}
Vaseghi, S.~V. and Frayling-Cork, R. (1992).
\newblock Restoration of old gramophone recordings.
\newblock {\em Journal of the Audio Engineering Society}.

\bibitem[Vincent et~al., 2018]{vincent_audio_2018}
Vincent, E., Virtanen, T., and Gannot, S., editors (2018).
\newblock {\em Audio source separation and speech enhancement}.
\newblock John wiley \& sons edition.

\bibitem[Wang and Wang, 2020]{wang_time-frequency_2020}
Wang, H. and Wang, D. (2020).
\newblock Time-frequency loss for {CNN} based speech super-resolution.
\newblock In {\em International {Conference} on {Acoustics}, {Speech} and {Signal} {Processing}}.

\bibitem[Wang et~al., 2015]{wang_speech_2015}
Wang, Y., Zhao, S., Liu, W., Li, M., and Kuang, J. (2015).
\newblock Speech bandwidth expansion based on deep neural networks.
\newblock In {\em Interspeech}.

\bibitem[Yoshida and Abe, 1994]{yoshida_algorithm_1994}
Yoshida, Y. and Abe, M. (1994).
\newblock An algorithm to reconstruct wideband speech from narrowband speech based on codebook mapping.
\newblock In {\em International Conference on Spoken Language Processing}.

\end{thebibliography}

\end{document}